\documentclass[reprint,aps,prl,twocolumn,superscriptaddress]{revtex4-1}
\usepackage{graphicx}
\usepackage{hyperref}

\def\beq{\begin{equation}}
\def\eeq{\end{equation}}
\def\bea{\begin{eqnarray}}
\def\eea{\end{eqnarray}}

\begin{document}

\title{Why is High Energy Physics Lorentz Invariant?}

\author{Niayesh Afshordi}
\affiliation{Perimeter Institute
for Theoretical Physics\\ 31 Caroline Street North, Waterloo, ON, N2L 2Y5, Canada}
\affiliation{Department of Physics and Astronomy, University of Waterloo\\ 200 University Avenue West, Waterloo, ON, N2L 3G1, Canada }

\date{\today}

\begin{abstract}

Despite the tremendous empirical success of equivalence principle, there are several theoretical motivations for existence of a preferred reference frame (or aether) in a consistent theory of quantum gravity. However, if quantum gravity had a preferred reference frame, why would high energy processes enjoy such a high degree of Lorentz symmetry? While this is often considered as an argument against aether, here I provide three independent arguments for why perturbative unitarity (or weak coupling) of the Lorentz-violating effective field theories put stringent constraints on possible observable violations of Lorentz symmetry at high energies. In particular, the interaction with the scalar graviton in a consistent low-energy theory of gravity and a (radiatively {\it and} dynamically) stable cosmological framework, leads to these constraints. The violation (quantified by the relative difference in maximum speed of propagation) is limited to $\lesssim 10^{-10} E({\rm eV})^{-4}$ (superseding all current empirical bounds), or the theory will be strongly coupled beyond meV scale. The latter happens in  extended Horava-Lifshitz gravities, as a result of a previously ignored quantum anomaly.  Finally, given that all {\it cosmologically viable} theories with significant Lorentz violation appear to be strongly coupled beyond meV scale, we conjecture that, similar to color confinement in QCD, or Vainshetin screening for massive gravity, high energy theories (that interact with gravity) are shielded from Lorentz violation (at least, up to the scale where gravity is UV-completed). In contrast, microwave or radio photons, cosmic background neutrinos, or gravitational waves may provide more promising candidates for discovery of  violations of Lorentz symmetry. 

\end{abstract}
\maketitle

\section{Introduction}

{\it Equivalence principle}, which lays at the heart of Einstein's special and general theories of relativity, is easily one of the most profound and arguably the best-tested tenet of modern physics \cite{Will:2005va}. In particular, high energy processes appear to enjoy a remarkable degree of Lorentz symmetry \cite{Kostelecky:2008ts, Liberati:2013xla}. 

Nevertheless, many theoretical motivations for a fundamental preferred frame (or {\it aether}), which, in principle, would violate Lorentz symmetry exist. These include attempts at a renormalizable or UV-complete theory of quantum gravity (e.g., \cite{Horava:2009uw}), possible solutions to the old \cite{Afshordi:2008xu,Kamiab:2011am,Aslanbeigi:2011si,Narimani:2014zha} and new cosmological constant problems \cite{PrescodWeinstein:2009mp}, early universe cosmology \cite{Mukohyama:2009gg,Magueijo:2008pm}, or microstates of {\it firewalls} \cite{Saravani:2012is}.

Is it possible for nature to be such an observant adherent to a symmetry, an it not be a fundamental aspect of nature? It is often argued that such a setting would be highly unstable to radiative (or quantum loop) corrections, and requires extreme fine-tuning (e.g., \cite{Collins:2004bp,Polchinski:2011za,LopezNacir:2011mt}). While this is widely taken as evidence against aether theories, there are already indications that such a conclusion might be premature, e.g.,
\begin{enumerate}
\item One may imagine custodial symmetries (such as supersymmetry) that forbid generation of  highly constrained dimension 4 Lorentz-violating operators \cite{GrootNibbelink:2004za}.

\item Since the main motivation for a fundamental violation of Lorentz symmetry comes from quantum gravity, it is possible (albeit not terribly satisfying) to imagine that Lorentz violation is constrained to the gravity sector, as long as the scale of UV completion for gravity, $\Lambda_g$ is well below the Planck scale, $M_p$. In this case, Lorentz violation in the matter sector  generated by radiative corrections is limited to $\lesssim \Lambda^2_g/M^2_p$  \cite{Pospelov:2010mp}. 

\item Concrete arguments about radiative corrections to Lorentz violating operators in the IR rely on weak coupling of matter fields to a (semi)-classical Lorentz violating background, or aether. One loses perturbative control if either couplings to matter or aether become strong, and non-perturbative methods have been invoked to construct an efficient emergence of Lorentz symmetry in the IR (e.g., \cite{Anber:2011xf, Bednik:2013nxa}). 
\end{enumerate}

In this {\it letter}, we seek to convince the reader that the latter is {\it inevitable} in any (known) cosmologically viable theory that violates Lorentz invariance. More precisely, the requirements of: 
\begin{enumerate} \item Weak coupling to a semi-classical cosmological reference frame, {\it and}  \item Consistent coupling to general relativity (GR),
\end{enumerate}
 severely limits any possible violation of Lorentz symmetry (well below any current empirical bound) for energies above $\sim$ meV, up to the scale where GR is UV completed. Therefore, the mere dismissal of a Lorentz-violating UV-completion of our theories based on perturbative (or radiative loop) corrections, is not justified.

\section{setup: a preferred time}

Let us start with a simple Lorentz-violating Lagrangian for $\phi$ field:
\beq
{\cal L}_{\phi} = \frac{1}{2}\partial^\mu\phi\partial_\mu\phi+\epsilon \dot{\phi}^2,\label{free_phi}
\eeq
which yields a modified speed of propagation:
\beq
c_{\phi}^2=(1+2\epsilon)^{-1}, {\rm ~or~} c_{\phi} = 1-\epsilon +{\cal O}(\epsilon^2).
\eeq 

While this is not the most general possible Lorentz violation, an $\epsilon$ that may depend on energy and type of particle is a generic expectation from coupling standard model fields to aether. Depending on the assumed energy dependence and particle type, upper bounds on $|\epsilon|$ range from $10^{-16}$ to $10^{-32}$ for particles with $ E \gtrsim  $ GeV \cite{Liberati:2013xla,Stecker:2013jfa}. 

While Eq. (\ref{free_phi}) describes a free theory, coupling to gravity will introduce interactions with $\phi$. This can be done most easily in a covariant framework, which separates the scalar and tensor polarizations of the graviton. Despite the violation of general covariance in (\ref{free_phi}), it can be recovered via the Stuckelberg trick, which is equivalent to considering the time in the preferred frame of (\ref{free_phi}) as a scalar field, $\tau$ in a general coordinate system:
\beq
{\cal L}_{\phi} = \frac{1}{2}\partial^\mu\phi\partial_\mu\phi + \epsilon \left(\partial^\mu \tau \partial_\mu \phi\right)^2.\label{cov_phi}
\eeq
Note that in the preferred reference frame of Eq. (\ref{free_phi}), the dynamics of $\tau$ is carried by the scalar polarization of the graviton. The covariantization procedure is necessary to ensure that gravitational constraint equations (or Bianchi identities) are satisfied, when the theory is coupled to gravity. Otherwise, coupling gravity to a non-covariant theory generally leads to inconsistencies. Now, the full interacting theory can be written in a covariant form:
\beq
{\cal L} = {\cal L}_{\phi}+ {\cal L}_{\tau}(\tau,\partial \tau, \partial\partial\tau, \dots)+\frac{1}{2}M^2_pR\{g_{\mu\nu}\},
\eeq
where we assumed Einstein-Hilbert action for the spin-2 gravitons, $g_{\mu\nu}$. The reason for the introduction of ${\cal L}_\tau$ is that ${\cal L}_\phi$ only includes interactions for $\tau$, and thus a separate kinetic term for $\tau$ is necessary to have perturbative control of the quantum theory. 

\section{1st Argument: Aether as a scalar Effective Field Theory}

In a cosmological setting, and in the absence of inhomogeneties, symmetries of the FRW space-time fix the preferred frame to be the FRW comoving frame (or the cosmic microwave background rest frame). Therefore, in general we can write: 
\beq
\tau({\bf x},t) = t + \alpha \psi({\bf x},t), \label{psi_def}
\eeq
where $t$ is the cosmic proper time, and $ \psi({\bf x},t)$ vanishes in the limit that  inhomogeneties (or deviations from FRW space-time) go to zero. $\alpha$ is chosen so that ${\cal L}_\tau$ has canonical normalization to quadratic order in $\psi$, at the present epoch. Moreover, to prevent the Jeans instability of aether, the depth of the gravitational potential needs to be shallower than the square of the effective sound speed for the aether, $c^2_\psi$. Given that the Galactic escape velocity at the solar circle is roughly $600$ km/s, this implies:
\beq
c_\psi \gtrsim 2 \times 10^{-3}.
\eeq
For example, this bound precludes the ghost condensate model \cite{ArkaniHamed:2003uy} with $c_\psi =0$, which tends to form caustics and break down (as an effective field theory) upon collapse into the Galactic potential. 

Now, the action for $\psi$ takes the form:
\beq
\delta{\cal L}_\tau \simeq \frac{1}{2} \left[\dot{\psi}^2-c_\psi^2 (\nabla \psi)^2\right] +\frac{\alpha^2}{2}(1-c_\psi^2)\left(\partial_\mu\psi\partial^\mu\psi\right)^2 + \dots ,\label{aether_eft}
\eeq
where we used Eq. (\ref{cov_phi}) (replacing $\phi$ with $\psi$) to covariantize the quadratic action for $\psi$. In other words, covariantization requires a minimal self-interaction for fields with a condensate with speed of sound $c_\psi \neq 1$ \footnote{we have ignored a cubic interaction term, which will be subdominant at high energies}. 

We can estimate the energy cut-off of the theory, $\Lambda_\psi$, by comparing the expectation value of the self-interaction term with that of the free Hamiltonian. Using the fact that that:
\beq
\langle \psi^2 \rangle_{\rm free} = \int \frac{d^3k}{2(2\pi)^3 \omega(k)} = \frac{\Lambda^2_\psi}{2(2\pi)^2c^3_\psi},
\eeq
which fixes $\Lambda_\psi$:
\beq
\frac{\Lambda^4_\psi}{16\pi^2c^3_\psi} \sim \frac{3}{2}\alpha^2|1-c^2_\psi|\left[\Lambda^4_\psi(1+c_\psi^2)\over (4\pi)^2 c^5_\psi\right]^2,
\eeq
i.e. the aether effective field theory (\ref{aether_eft}) is valid for:
\beq
E < \Lambda_\psi \sim \left[32\pi^2 c_\psi^7\over 3\alpha^2|1-c^2_\psi |(1+c^2_\psi)^2 \right]^{1/4}.\label{psi_cutoff}
\eeq

Interestingly, we can put a lower limit on $\alpha$, by putting aether in a  cosmological framework. We first notice that, for the background (\ref{psi_def}), the arguments of ${\cal L}_\tau$, take the following form:
\beq
{\cal L}_\tau(\tau,\partial \tau, \partial\partial\tau, \dots) = {\cal L}_\tau\left(t+\alpha\psi, \delta^0_\mu+\alpha \partial_\mu\psi, \alpha^2\partial_\mu\partial_\nu\psi, \dots\right),\label{l_tau_diff}
\eeq
Defining:
\beq
X \equiv \frac{1}{2}g^{\mu\nu} \partial^\nu\tau\partial_\mu\tau = \frac{1}{2} + \alpha \dot{\psi}+\frac{1}{2}\alpha^2\left(\partial_\mu\psi\partial^\mu\psi\right),
\eeq
we first notice that the only dependence of {\it background} Lagrangian ${\cal L}^{(0)}_\tau$ (i.e. with $\psi =0$) on metric $g_{\mu\nu}$ is through its dependence on $X$. Therefore,
\beq
T_{\mu\nu}^{(0)} = \frac{2}{\sqrt{-g}}\frac{\partial (\sqrt{-g}{\cal L}^{(0)}_\tau)}{\partial g^{\mu\nu}} =  \frac{\partial {\cal L}_\tau}{\partial X} \delta^0_\mu\delta^0_\nu - {\cal L}_\tau g_{\mu\nu},\label{T_back}
\eeq
or consequently, the sum of aether background pressure and density is given by:
\beq
\rho^{(0)}+p^{(0)} = \frac{\partial {\cal L}_\tau}{\partial X}. 
\eeq

Incidentally, the only dependence of  ${\cal L}_\tau$ on $(\nabla\psi)^2$ (Eq. \ref{l_tau_diff}) is also through its dependence on $X$. Therefore, using Eq. (\ref{aether_eft}), we get:
\beq
-\frac{1}{2}c^2_\psi = \frac{\partial {\cal L}_\tau}{\partial X} \frac{\partial X}{\partial (\nabla\psi)^2}= -\frac{1}{2}(\rho^{(0)}+p^{(0)})\alpha^2.
\eeq
Given the cosmological constraints on dark energy density and pressure, this yields:
\beq
c^2_\psi \alpha^{-2} = \rho^{(0)}+p^{(0)} < 3 \Omega_\Lambda |1+w|  M^2_p H^2 < (1.3~{\rm meV})^4,
\eeq
where $w$ is the dark energy equation of state, which is constrained to $|1+w| <0.09$ at $\sim 95\%$ confidence level \cite{Ade:2015xua}, while $\Omega_{\Lambda}$, $M_p$, and $H$ are the dark energy density parameter, reduced Planck mass, and present-day Hubble constants respectively.  Plugging this into Eq. (\ref{psi_cutoff}) yields:
\beq
E < \Lambda_\psi \lesssim (4~ {\rm meV}) \left[c_\psi^5\over |1-c^2_\psi |(1+c^2_\psi)^2 \right]^{1/4}.\label{psi_cutoff_cosmo}
\eeq

We thus find that the field theory for aether has a very low cut-off, unless aether perturbations propagate with a speed very close to the speed of light, $c_\psi \simeq 1$ (i.e. they are almost Lorentz-invariant).  We'll next see that this puts severe bounds on Lorentz violation via coupling to aether, in an effective field theory framework.  
 
Now turning to the $\phi-\psi$ (i.e. matter-aether) interactions, the 4th order interaction term in Eq. (\ref{cov_phi}) takes the form:
\beq
\delta {\cal L}^{(4)}_{\psi,\phi}  =\epsilon \alpha^2 ( \partial^\mu \psi \partial_\mu \phi)^2. 
\eeq

The $\phi$ effective action loses perturbative control, when the expectation value for interaction term, becomes comparable to the (free) kinetic term, i.e. when 
\beq
\epsilon \alpha^2\langle \partial\psi \partial\psi \rangle \sim 1.
\eeq
Computing the expectation value yields a maximum cut-off for the matter+aether effective field theory of, $\Lambda_\phi$:
\beq
|1-c_\phi| \alpha^2 \left[\Lambda^4_\phi(1+c_\psi^2)\over (4\pi)^2 c^5_\psi\right] \sim 1,
\eeq
or
\beq
\Lambda_\phi \sim \left[16\pi^2 c_\psi^5\over \alpha^2|1-c_\phi|(1+c_\psi^2)\right]^{1/4}.
\eeq

Now, for the effective field theory to be under control, both $\phi$ and $\psi$ need to be weakly coupled, implying that the particle energies should be less than both $\Lambda_\phi$ and $\Lambda_\psi$. These cut-offs can be combined to give:
\beq
E^{-4} \gtrsim \Lambda^{-4}_\phi+\Lambda^{-4}_\psi, 
\eeq
which yields:
\bea
|1-c_\phi | &\lesssim& \frac{(4\pi)^2c^5_\psi}{\alpha^2(1+c^2_\psi)E^4}-\frac{3}{2}|c^{-2}_\psi-c^2_\psi|  \nonumber\\
&\lesssim&   \frac{c^3_\psi}{(1+c^2_\psi)}\left(E \over 5~{\rm meV}\right)^{-4}-\frac{3}{2}|c^{-2}_\psi-c^2_\psi|. 
\eea
We see that for $E \gtrsim 4$ meV, the right hand side is maximized at $c_\psi =1$, which yields:
\beq
\epsilon = |1-c_\phi| \lesssim \left(E \over 4~{\rm meV}\right)^{-4}, {\rm ~ for ~} E\gtrsim 4{\rm ~ meV}.\label{eps_bound}
\eeq
We note that the lowest energies at which Lorentz-violation (in the form of difference in the maxim speed of propagation of different particle species) has been tested is at electron mass $ m_e = 511$ keV, for which Eq.  (\ref{eps_bound}) implies: $\epsilon \lesssim 10^{-33}$, which is far below any experimental limit \cite{Liberati:2013xla,Stecker:2013jfa}, and drops very steeply at higher energies. While in principle low-energy neutrinos can show larger Lorentz-violation,  solar or reactor neutrinos,  that are probed in current experiments, have energies $\sim $ MeV, and thus have similarly low upper limits on their Lorentz-violation (assuming cosmology and weak coupling). 

We should note that simpler versions of this argument have already appeared in the literature (e.g., \cite{Dvali:2007ks,Afshordi:2012py}), and all we have done so far, is to make them more systematic for Lorentz violation due to a cosmological preferred time.  What is missing from these arguments is the possibility of additional (self-)interaction terms that cancel the ones we have considered above, due to new symmetries in the aether sector. We shall address this possibility next.
 
\section{2nd Argument: Anomaly in Einstein-Aether and Horava-Lifshitz gravities}
  
The only way to avoid the above constraints is to introduce a richer structure to the interactions, which are protected by possible symmetries, and could lead to cancelation of UV divergences. To our knowledge, the only such known symmetry is field reparametrization (or {\it Khronon}) symmetry, i.e. ${\cal L}_\tau$ should be invariant under $\tau \rightarrow f(\tau)$. This is equivalent to time reparametrization in the so-called healthy extensions of Horava-Lifshitz gravity \cite{Blas:2009qj}, as well as hypersurface orthogonal Einstein-Aether theories \cite{Jacobson:2010mx}\footnote{Hypersurface orthogonality condition is necessary to for the existence of a global reference frame.}, where the action is constructed out of the unit vector $u^\mu$:
\beq
u^\mu \equiv \frac{\partial^\mu \tau}{\sqrt{\partial^\alpha\tau\partial_\alpha\tau}}.
\eeq 
What this implies is that, if the Lorentz violation is generated through coupling to $u^\mu$, then $\epsilon \propto (\partial \tau)^{-1}$ in Eq. \ref{cov_phi}, which regulates the UV divergence of the quartic coupling. 

The scalar  field in both theories takes the following covariant form at low energies \cite{Jacobson:2010mx}:
\beq
{\cal L}_{\tau, EA} = \frac{M^2}{c^3_\psi} \left[ c^{-2}_\psi \dot{u}^\mu\dot{u}_\mu - (\nabla_\mu u^\mu)^2\right],\label{EA_action}
\eeq 
where $\dot{} \equiv u^\alpha \nabla_\alpha$, and $M$ is roughly the energy cut-off of the effective field theory based on power-counting of the minimal covariant interactions in (\ref{EA_action}) \footnote{There is a 3rd parameter in this class of theories that can be set to zero by  a disformal metric transformation  \cite{Jacobson:2010mx}. This would lead to additional couplings of matter to $u^\mu$ in terms of the new metric, which simply renormalize the bare Lorentz-violating interactions in the aether frame. }. This is in apparent contradiction  with our result in (\ref{psi_cutoff_cosmo}), as the cosmological density/pressure is $\sim M^2H^2/c^3_\psi$, and thus the cut-off $M$ could be close to Planck mass (assuming $c_\psi \gtrsim 1$) without violating cosmological bounds. However, we next argue the true cut-off of the quantum theory should be much lower than $M$.  

To see this, we note that the quantum theory is described by the path integral:
\beq
{\cal Z} \equiv \int Dg_{\mu\nu} D\tau ~ {\cal M}[g_{\mu\nu},\tau] e^{i{\cal S}_{\rm cl}[g_{\mu\nu},\tau]}. 
\eeq
While the classical action, ${\cal S}_{\rm cl}[g_{\mu\nu},\tau] =  {\cal S}_{GR}+\int d^4x \sqrt{-g} {\cal L}_{\tau, EA}$, for Einstein-Aether theories is invariant under field reparametrization $\tau \rightarrow f(\tau)$, quantum correction will break the  symmetry unless the measure $D\tau~{\cal M}[g_{\mu\nu},\tau] $ is also invariant. Therefore, we should seek a measure ${\cal M}$ that is invariant under field reparametrization. 

Let us start by considering two space-time points, and requiring that $ d\tau_1 d\tau_2 {\cal M}(\tau_1,\tau_2)$ is invariant under linear field reparametrizations: $\tau_i \rightarrow A\tau_i+B$. This requirement uniquely fixes the measure:  $ {\cal M}(\tau_1,\tau_2) \propto (\tau_1-\tau_2)^{-2}$.  Extending this to continuum space-time uniquely fixes a covariant measure (up to 2nd or higher order derivatives):
\beq
 D\tau~{\cal M}[g_{\mu\nu},\tau] = \prod_x \frac{d\tau_x}{\sqrt{|\partial^\mu\tau\partial_\mu\tau|}},
\eeq
which is now invariant under general field reparametrizations. In other words, the quantum effective action now has an imaginary contribution from the measure ${\cal M}$:
\beq
{\cal S}_{\rm q}[g_{\mu\nu},\tau] = {\cal S}_{\rm cl}[g_{\mu\nu},\tau] +\frac{i\Lambda^4_\psi}{2c^3_\psi} \int d^4x \sqrt{-g} \ln|\partial^\mu\tau\partial_\mu\tau|,
\eeq
where we have taken $\Lambda_\psi$ and $\Lambda_\psi/c_\psi$ to be the cut-off's in energy and momentum for the effective field theory. Now, the quantum effective Lagrangian for the Einstein-Aether theory (\ref{EA_action}), to quadratic order in $\psi$ field (\ref{psi_def}) on sub-horizon scales is given by:
\bea
&&{\cal S}_{\rm q} = \frac{\alpha^2}{c^3_{\psi}}\int \frac{d\omega d^3k}{(2\pi)^4} \times \nonumber\\ &&  \left[ \left(\frac{M^2k^2}{c^2_\psi} - \frac{i\Lambda^4_\psi}{2}\right) \omega^2- \left(M^2k^2+\frac{i\Lambda^4_\psi}{2} \right)k^2\right] |\psi_{\omega, {\bf k}}|^2, \nonumber\\&&
\eea
in Fourier space. The Feynman propagator, defined as $\langle T \psi(x) \psi(y)\rangle$ for the free theory, has a complex pole in the momentum space at:
\beq
\omega^2 = k^2c^2_\psi \left(  2M^2k^2+i\Lambda^4_\psi  \over 2M^2k^2 - i c^2_\psi \Lambda^4_\psi \right),
\eeq
which points to a {\it dynamical} instability, with the maximum rate of:
\beq 
\max \Im \omega \sim \frac{c_\psi(1+c^2_\psi)^{1/2} \Lambda^2_\psi}{M}.
\eeq
For the theory to be stable/unitary on cosmological scales, we need the instability rate to be less than the Hubble rate, $\Im \omega < H$ , which puts an upper limit on the UV cut-off of the the theory, $\Lambda_\psi$:
\beq
\Lambda^4_\psi \lesssim \frac{H^2 M^2}{c^2_\psi (1+c^2_\psi)} \lesssim  \frac{(1 ~{\rm meV})^4}{c_\psi+c^{-1}_\psi}  \lesssim  (1 ~{\rm meV})^4,\label{EA_cutoff}
\eeq
where we used the fact that the aether density/pressure $\sim M^2 H^2/c^3_\psi$ should be less than the critical density $3 M^2_p H^2 \sim (1 ~{\rm meV})^4$. 

Therefore, we see that the real UV cut-off of the Einstein-Aether theory should be less than $\sim$ meV, far below its naive cut-off of $M$ (which could have been comparable to Planck energy). We thus conclude that direct coupling to Einstein-Aether theories cannot violate Lorentz symmetry in high energy processes, at least in a perturbative setting. The same applies to the ``healthy extensions'' of Horava-Lifshitz gravity, which are equivalent to hypersurface-orthogonal Einstein-Aether theories at low energies \cite{Jacobson:2010mx}.  It is worth noting that a similar meV cut-off applies to Lorentz-violating theories of massive gravity \cite{Dubovsky:2004sg,Comelli:2013txa}, with a Hubble scale mass. 


\section{3rd Argument: 2nd cosmological constant problem}

Now, let us compute the contribution of the $\phi$ field to the vacuum energy-momentum tensor.  Since $\phi$ is covariantly coupled to the metric:
\beq
\tilde{g}_{\mu\nu}  = g_{\mu\nu}+2\epsilon \partial_\mu\tau\partial_\nu\tau,
\eeq
then vacuum expectation value of the energy-momentum tensor for $\phi$ is expected to be:
\beq
\langle T_{\mu\nu} \rangle \sim \Lambda^4_\phi \tilde{g}_{\mu\nu} =  \Lambda^4_\phi \left( g_{\mu\nu}+2\epsilon \partial_\mu\tau\partial_\nu\tau\right), \label{vac_em}
\eeq
assuming that UV physics for $\phi$ respects the Lorentz symmetry of $\tilde{g}_{\mu\nu}$. While the quartic divergence of the first term in Eq. (\ref{vac_em}) as $\Lambda_\phi \rightarrow \infty$ represents the old cosmological constant problem, the presence of the second term makes the vacuum energy a dynamical entity, whose divergence could be potentially more dangerous. In particular, the second term contributes to $\rho+p$ of aether, which without assumption of fine-tuning implies:
\bea
\rho +p = 2 \Lambda^4_\phi  \epsilon   < 3 \Omega_\Lambda |1+w|  M^2_p H^2 < (1.7~{\rm meV})^4, \nonumber\\
\Rightarrow \epsilon = |1-c_\phi| \lesssim \left(\Lambda_\phi \over 1.4~{\rm meV}\right)^{-4} <\left(E \over 1.4~{\rm meV}\right)^{-4},\nonumber\\
\eea 
similar to what we found in Eq. (\ref{eps_bound}).

\section{Discussion}

The systematic (theoretical and empirical) exploration of possible Lorentz-violating extensions of standard model has been a very fruitful field of research over the past couple of decades \cite{Colladay:1998fq,Kostelecky:2008ts, Liberati:2013xla}. What has been lacking from these studies have been a consistent coupling to gravity, which drives the universe on cosmological scales. We have provided considerable evidence in this {\it letter} that including gravity in a cosmological framework {\it inevitably} leads to strong coupling, and loss of quantum perturbative control for any Lorentz-violating theory of high energy physics $>$ meV. 

We should note that strong coupling is {\it not} a physical problem for a theory, but rather for the perturbative calculations (e.g. at tree or finite loop order). Two well-known examples of strong coupling in physics are color confinement in QCD and Vainshetin screening in massive gravity theories \cite{Vainshtein:1972sx,Babichev:2013usa}. In both examples, further non-perturbative analyses (or insights) have shown that the strongly coupled degree of freedom (i.e. gluon in QCD or scalar graviton in massive gravity) effectively decouples from the rest of theory. 
This leads us to conjecture that similarly, aether must effectively {\it decouple} from high energy physics beyond meV scales, up to the scale where GR is UV completed (and presumably does not obey the covariant description used here). So, maybe precision confirmations of Lorentz symmetry at high energies should not have been so surprising.  


Now, where {\it should} we look for possible violations of Lorentz symmetry? The lesson from this analysis is that low energy particles, with $E \lesssim $ meV, e.g., microwave or radio photons, cosmic background neutrinos, or gravitational waves can weakly couple to cosmological aether, and thus may provide more promising candidates for discovery of  violations of Lorentz symmetry.

{\it Acknowledgement}: I would like to thank Ted Jacobson, Oriol Pujolas, and especially Maxim Pospelov for useful discussions. This research is supported by the Perimeter Institute for Theoretical Physics and the University of Waterloo. Research at the Perimeter Institute is supported in part by the Government of Canada through Industry Canada, and by the Province of Ontario through the Ministry of Research and Information (MRI) 

\bibliography{Lorentz.bib}

\end{document}